\documentclass[12pt]{article}
\usepackage{amssymb}
\usepackage{epsfig}

\catcode`\@=11
\def\numberbysection{\@addtoreset{equation}{section}
        \def\theequation{\thesection.\arabic{equation}}}
\def\half{\frac{1}{2}}
\def\beq{\begin{equation}}
\def\eeq{\end{equation}}
\numberbysection
\begin{document}
\begin{titlepage}
\begin{center}
\hfill DFF 1/9/96  \\
\vskip 1.in
{\Large \bf (2+1)-Gravity Solutions with Spinning Particles}
\vskip 0.5in
M. Ciafaloni and P. Valtancoli
\\[.2in]
{\em Dipartimento di Fisica dell' Universita', Firenze \\
and INFN, Sezione di Firenze (Italy)}
\end{center}
\vskip .5in
\begin{abstract}
We derive, in 2+1 dimensions, classical solutions for metric and motion 
of two or more spinning particles, in the conformal Coulomb gauge introduced
previously. The solutions are exact in the $N$-body static case, and
are perturbative in the particles' velocities in the dynamic two-body
case. A natural boundary for the existence of our gauge choice is provided
by some ``CTC horizons'' encircling the particles, within which closed timelike
curves occur.
\end{abstract}
\medskip
\end{titlepage}
\pagenumbering{arabic}
\section{Introduction}

Interest in the classical solutions of $2+1$-Gravity
\cite{a1}-\cite{a10} has recently revived \cite{a11}-\cite{a13} because of
the discovery of exact moving particle solutions \cite{a12}-\cite{a13}
in a regular gauge of conformal type \cite{a11}. Simplifying features of such a
gauge are the instantaneous propagation ( which makes the ADM decomposition
of space-time explicit and particularly simple ) and the conformal
factor of Liouville type (which can be exactly found at least in the two-body
case).

We have already provided in Ref. \cite{a12}, hereafter referred to as [BCV], the
main results for the case of $N$ moving spinless particles. The
purpose of the present paper is to extend the BCV gauge choice \cite{a11}
to spinning particles and to provide solutions for the metric and the
motion in some particular cases.

Localized spin $s$, in $2+1$ dimensions \cite{a6}, is characterized by the
fact that a Minkowskian frame set up in the neighbourhood of the particle
has a multivalued time, which is shifted by the amount $\delta T = -
s$, when turning around it in a closed loop. A consequence of this
jump ( which is backwards in time for a proper loop orientation ) is that
there are closed timelike curves ( CTC's ) \cite{a14}
around the particles at a distance smaller
than some critical radius $R_0 \sim O(s)$.

This feature suggests that the single-valued time of our gauge, which
is syncronized in  a global way, cannot be pushed too close to the particles
themselves. Indeed we shall find that there are `` CTC horizons'' around
the particles of radii $R_i \sim s_i$ which cannot be covered by our gauge
choice \cite{a15}. Nevertheless, we will be able to describe the
motion of the particles themselves on the basis of our ``external
solutions'' to the metric and to the DJH matching conditions.

Technically, the existence of the time shifts mentioned before
modifies the number of ``apparent singularities'' which appear in the
Riemann-Hilbert problem \cite{a16} for the analytic function providing the
mapping to Minkowskian coordinates. Such singularities are not branch points
of the mapping function, but nevertheless appear as poles in its
Schwarzian derivative.

While for $N$ spinless particles there are $2N-1$ singularities ( $N$
for the particles, $1$ at infinity and $N-2$ apparent singularities ),
in the spinning case there are $3N-1$, corresponding to one more apparent
singularity per particle. This means that explicit exact solutions are more
difficult to find.

In the spinless case we found exact solutions for the two-body problem
with any speed ( $3$ singularities ) and for $N$ bodies with small speed.
In the spinning case we find here an exact solution only for the
static ( $N$-body ) case, and we discuss the two-body problem, which 
corresponds to five singularities,  for the case of small speed only.

The outline of the paper is as follows. In Sec. 2 we recall the general
features of our method in the conformal Coulomb gauge in both first-order
and ADM \cite{a17} formalisms. In particular, we show how the metric
can be found once the mapping function $f(z,t)$ and the meromorphic
function $N(z,t)$ are given. In Sec. $3$ we give an exact solution for
$f$ and $N$, in the
case of spinning particles at rest, characterized by the fact that $N$ has
double poles at the particle sites, with residues proportional to the
spins. We show that such double poles, related to an energy-momentum
density of ${\delta}'$-type, are at the origin of the time shifts, of
the apparent singularities, and of the $CTC$ horizons close to the particles.
In Sec. $4$ we discuss the two-body problem, corresponding to $5$
singularities, at both first-order and second-order in the velocities. The 
second-order solution corresponds to the non-relativistic limit and is
of particular interest, even in the spinless case.

Our results and conclusions are summarized in Sec. $5$, and some technical
details are contained in Appendices $A$ and $B$.

\section{General features and gauge choice} 

{\bf 2.1 From Minkowskian to single-valued coordinates }

In [BCV] we have proposed a non-perturbative solution for the metric
and the motion of $N$ interacting spinless particles in (2+1)-gravity,
based on the introduction of a new gauge choice which yields an instantaneous
propagation of the gravitational force.

Our gauge choice is better understood in the first-order formalism
which naturally incorporates the flatness property of $(2+1)$ space-time 
outside the sources. This feature allows to choose a global Minkowskian 
reference system $X^a \equiv (T, Z, \overline{Z} )$, which however is in 
general multivalued, due to the localized curvature at the particle
sources. In order to have well-defined coordinates, cuts should be
introduced along tails departing from each particle, and a Lorentz
transformation should relate the values of $dX^a$'s along the cuts, so
that the line element $ ds^2 = \eta_{ab} dX^a dX^b $ is left  
single-valued. 

The crucial point of our method is to build a
representation of the $X^a$'s starting from a regular coordinate
system $x^\mu = ( t , z, \overline{z} )$, as follows:

\beq dX^a = E^a_\mu dx^\mu = E^a_0 dt + E^a_z dz + E^a_{\overline z} d
\overline{z} \eeq

Here the dreibein $E^a_\mu$ is multivalued and satisfies the
integrability condition :

\beq \partial_{[\mu} E^a_{\nu ]} \ = \ 0 \eeq

which implies a locally vanishing spin connection, outside the
particle tails \cite{a9}.

 Let us choose to work in a Coulomb gauge :

\beq \partial \cdot E^a = \partial_z E^a_{\overline z} +
\partial_{\overline z} E^a_z = 0 \eeq

which, together with the equations of motion (2.2), implies

\beq \partial_z E^a_{\overline z} = \partial_{\overline z} E^a_z = 0 ,
\eeq

so that $ E^a_z ( E^a_{\overline z} ) $ is analytic ( antianalytic ).

Multiplying (2.4) by $E^a_z$ we also get $\partial_{\overline z}
g_{zz} = 0$; we choose to impose the conformal condition $ g_{zz} =
g_{{\overline z}{\overline z}} = 0 $ in order to avoid arbitrary analytic 
functions as components of the metric. Hence we can parametrize $E^a_z$, 
$E^a_{\overline z}$ in terms of null-vectors:

\beq E^a_z = N W^a  \ , \ \ \ E^a_{\overline z} = {\overline N}
{\widetilde W}^a \eeq

where $ W^2 = {\tilde W}^2 = 0 $, and we can assume $N(z,t)$ to be a
single-valued meromorphic function ( with poles at $z = \xi_i$, as we
shall see ).

We have to build $W^a$, ${\tilde W}^a $ in order to represent the DJH \cite{a2}
matching conditions of the $X^a$ coordinates, around the particle
sites $z = \xi_i (t)$:

\beq  (dX^a)_I \rightarrow (dX^a)_{II} = {(L_i)}^a_b (dX^b)_I \ \ \
i = 1, 2, ..., N \eeq

where $ L_i = exp ( i J_a P_i^a ) $ ( $(iJ_a)_{bc} = \epsilon_{abc}$ )
denote the holonomies of the spin connection, which is treated here in
a global way, in order to avoid distributions.

The simplest realization of such $O(2,1)$ monodromies is given by a
spin $\half$ projective representation:

\beq f(z,t) \rightarrow \frac{ a_i f(z,t) + b_i}{ b^*_i f(z,t) + a^*_i
}, 
\ \ \ \ a_i = \cos \frac{m_i}{2} + i \ \gamma_i \ \sin \frac{m_i}{2} , \ \ \
b_i = - i \ \gamma_i \ \overline{V}_i \sin \frac{m_i}{2} \eeq

where the mapping function $f(z,t)$ is an analytic function with branch-cuts 
at $ z = \xi_i (t) $ and $V_i = P_i/ E_i$ ( $ \gamma_i = {( 1 -
{|V_i|}^2 ) }^{-\half} $ ) are the constant Minkowskian velocities.

Since the $W$ vectors must transform according to the adjoint representation
of $O(2,1)$, the natural choice, constructed out of the mapping function $f$ 
defined before, is to set :

\beq W^a = \frac{1}{f'}( \ f, \ 1, \ f^2 ) \ , \ \ \ \ {\widetilde W}^a =
\frac{1}{\overline f'}( \ {\overline f}, \ {\overline f}^2, \ 1 ) \eeq

which gives for the spatial component $g_{z{\overline z}}$ of the
metric tensor the expression

\beq - 2 g_{z{\overline z}} \ = \ e^{2\phi} \ = \ {|\frac{N}{f'}|}^2 ( 1 - {|
f |}^2 )^2 \eeq

in which we recognize the general solution of a Liouville-type
equation \cite{a18}.

We can now integrate (2.1) out of particle 1 :

\beq X^a = X^a_1 (t) + \int_{\xi_1}^z dz \ N \ W^a (z,t) +
\int_{{\overline \xi}_1}^{\overline z} d {\overline z} \ {\overline N}
{\widetilde W}^a ( {\overline z}, t ) \eeq

in terms of the parametrization $X^a_1 (t)$ of one Minkowskian
trajectory, which is left arbitrary.

The $X^a = X^a (x)$ mapping is at this point uniquely determined once
the solution to the monodromy problem (2.7) is found. Since the
coefficients $(a_i, b_i)$ are constants of motion, the monodromy
problem can be recast into a Riemann-Hilbert problem \cite{a16} for an
appropriate II order differential equation with Fuchsian
singularities, whose solutions are quoted in [BCV] for the spinless case.

For instance, in the two-body case, there are $3$ singularities, which can be
mapped to $\zeta_1 = 0$, $\zeta_2 = 1$, $\zeta_\infty  = \infty$,
where $\zeta = \frac{z - \xi_1}{\xi_2 - \xi_1}$, and the mapping
function is the ratio of two hypergeometric functions 

\begin{eqnarray}
f(\zeta) & = & \frac{\gamma_{12} {\overline V}_{12}}{\gamma_{12} - 1}
\zeta^{\mu_1} \frac{ {\tilde F}
\left(\half(1+\mu_\infty+\mu_1-\mu_2),\half(1-\mu_\infty+\mu_1-\mu_2),
1+\mu_1;\zeta\right)}{ {\tilde F}
\left(\half(1+\mu_\infty-\mu_1-\mu_2),\half(1-\mu_\infty-\mu_1-\mu_2),
1-\mu_1;\zeta\right)} \ , \nonumber \\ & & \nonumber \\
{\tilde F}(a,b,c;z) & \equiv & \frac{\Gamma (a) \Gamma (b)}{\Gamma (c)} 
F(a,b,c;z) \ ,  \ \ \ \gamma_{12} \equiv \frac{P_1 P_2}{m_1 m_2} 
\end{eqnarray}
whose difference of exponents are $\mu_i = m_i / 2 \pi $ ( i = 1,2 )and $\mu_\infty = 
{\cal M} / 2 \pi - 1$, where $\cal M$ is the total mass

\beq \cos(\frac{\cal M}{2}) = \cos (\frac{m_1}{2}) \cos (\frac{m_2}{2})
\ - \ \frac{P_1 \cdot P_2}{m_1 m_2} \ \sin (\frac{m_1}{2}) \sin 
(\frac{m_2}{2}) \eeq

In general, we can set 

\beq f = y_1 / y_2 \ , \ \ \ \ \ \ \ y''_i + q (\zeta) y_i = 0 , \eeq

where the potential

\beq 2 q(\zeta) = \{ f, \zeta \} = {\left( \frac{f''}{f'} \right)}' -
\half {\left( \frac{f''}{f'} \right)}^2 \eeq

is a meromorphic function with double and simple poles at the
singularities $\zeta = \zeta_i$.

It turns out that, for more than two particles, ``apparent singularities''
must be added into the differential equation in order to preserve the
constancy of the monodromy matrix with moveable singularities, as
firstly noticed by Fuchs \cite{a19}. Such singularities are zeros of
$f'$, rather
than branch points of $f$, and their position is related to the ones
of the particles in a generally complicated way, determined by the monodromies 
. The total number of singularities for $N$ particles is  $2N-1$, and
the mapping function was found for $N \ge 3$ in [BCV] in the limiting
case of small velocities.

In order to determine the metric completely, we derive (2.10) with respect
to time, and we obtain

\begin{eqnarray} E^a_0 \ & = & \ \partial_t X^a \ = \ \partial_t X^a_1 + \ 
\partial_t \left(
\int^z_{\xi_1} dz \ N \ W^a + \int^{\overline z}_{{\overline \xi}_1} \
d {\overline z} \ {\overline N} \ {\widetilde W}^a \right) \ = \ \nonumber \\
&  = &  c^a ( t ) + \
\int^z_{\xi_1} \ dz \ \partial_t (N W^a) + 
\int^{\overline z}_{{\overline \xi}_1}
\ d {\overline z} \ \partial_t ( {\overline N} {\widetilde W}^a ) 
\end{eqnarray} 

In terms of the vectors $E_0^a = \partial_t X^a$, $E^a_z = N W^a$,
$E^a_{\overline z} = {\overline N} {\widetilde W}^a $, the components of
the metric are given by :

\begin{eqnarray}
- 2 g_{z{\overline z}} \ & = & \ e^{2\phi} \ = \ {|N|}^2 ( - 2 W \cdot
{\tilde W} ) , \nonumber \\
g_{0z} \ & = & \ \half {\overline\beta} e^{2\phi} \ = \ N W_a E^a_0 , \ \
g_{0{\overline z}} \ = \ \half {\beta} e^{2\phi} \ = \ {\overline N} 
{\widetilde W}_a E^a_0 \nonumber \\
 g_{00} \ & = & \ \alpha^2 - {|\beta|}^2 e^{2\phi} = E^a_0 E^a_0, \ \
\ \ \ \ \alpha = V_a E^a_0 
\end{eqnarray}

so that the line element takes the form

\beq ds^2 \ = \ \alpha^2 dt^2 - e^{2 \phi} {| dz - \beta dt |}^2. \eeq

Here we have defined the unit vector 

\beq V^a = \frac{1}{1 - {|f|}^2} ( 1 + {|f|}^2, \ 2{\overline f}, \ 2 f )
\ = \ \epsilon^a_{bc} W^b {\widetilde W}^c {( W \cdot {\widetilde W})}^{-1} 
\eeq
which represents the normal with respect to the surface $X^a = X^a( t
, z, {\overline z})$, embedded at fixed time in the Minkowskian
space-time $ds^2 = \eta_{ab} dX^a dX^b $. The tangent plane is instead
generated by the vectors

\beq \partial_z X^a = N W^a , \ \ \ \ \ \ \partial_{\overline z} X^a =
{\overline N} {\widetilde W}^a . \eeq

We notice that it is not a priori warranted to have such a well
defined foliation of space-time in terms of surfaces at fixed
time. This probably requires the notion of a universal global time,
which is not valid for universes with closed time-like curves, as it
happens in the case of spinning sources. Hence we can anticipate that
we will have problems in defining our gauge globally for spinning
sources.

{\bf 2.2  The Einstein equations in the ADM formalism }

Quite similarly to what we have discussed now, the starting point of
the ADM formalism is to assume that space-time can be globally
decomposed as $\Sigma (t) \otimes R$, where $\Sigma(t)$ is 
 a set of space-like surfaces. The  
($2+1$)-dimensional metric is then split into ``space'' and ``time''
components:

\beq g_{00} = \alpha^2 - e^{2\phi} \beta {\overline \beta} , \ \ \ 
g_{0z} = \half {\overline\beta} e^{2\phi}, \ \ \  g_{0{\overline z}} = \half
\beta e^{2\phi} \ \ \ g_{z{\overline z}} = - \half e^{2\phi} \eeq
where $\alpha$ and $\beta$ provide the same parametrization as in Eq. (2.17).
The lapse function $\alpha$ and the shift functions $g_{0i}$ have the
meaning of Lagrange multipliers in the Hamiltonian formalism, since
their conjugate momenta are identically zero.

The ADM  space-time splitting  can be worked out from the Einstein-Hilbert
action by rewriting the scalar curvature $R^{(3)}$ into its spatial
part $R^{(2)}$, intrinsic to the surfaces $\Sigma (t)$, and an
extrinsic part, coming from the embedding , as follows

\beq S = - \half \int \sqrt{|g|} \ R^{(3)} = - \half \int \sqrt{|g|} \ \left[
R^{(2)} + {( Tr K )}^2 - Tr ( K^2 ) \right] \ d^3 x , \eeq

where the equivalence holds up to a boundary term. Here we have introduced
the extrinsic curvature tensor $K_{ij}$ , or second fundamental form
of the surface $\Sigma(t)$, given by :

\beq K_{ij} = \half \sqrt{ \frac{|g_{ij}|}{|g|} } \left( \nabla^{(2)}_i g_{0j}
+ \nabla^{(2)}_j g_{oi} - \partial_0 g_{ij} \right) \eeq

where we denote by $\nabla^{(2)}_i$ the covariant derivatives with
respect to the spatial part of the metric. The momenta $\Pi^{ij}$,
conjugate to $g_{ij}$ are proportional to $K^{ij} - g^{ij} K$ , which
therefore complete the canonical coordinate system of the Hamiltonian
formalism.

We can generate the ADM decomposition starting from the first-order
formalism foliation $X^a = X^a ( t, z , {\overline z})$. The Coulomb
gauge condition, imposed to fix  such a mapping, can be
directly related to the gauge condition of vanishing ``York time'' \cite{a20}

\beq g_{ij} \Pi^{ij} = K(z, {\overline z}, t) = K_{z{\overline z}} =
\frac{1}{2\alpha} ( \partial_z g_{0{\overline z}} +
\partial_{\overline z} g_{0z} - \partial_0 g_{z{\overline z}} )  = 0
\eeq

In fact, by rewriting this combination in terms of the dreibein we
get, by using eq. (2.2):

\beq E^a_0 \cdot ( \partial_z E^a_{\overline z} + \partial_{\overline
z} E^a_z ) + E^a_{\overline z} \cdot ( \partial_z E^a_0 - \partial_0 E^a_z )
+ E^a_z \cdot ( \partial_{\overline z} E^a_0 - \partial_0 E^a_{\overline z}
) = 0 \eeq

We thus see that our gauge choice is defined by the conditions 

\beq g_{zz} = g_{{\overline z}{\overline z}} = K = 0 \eeq

and thus corresponds to a conformal gauge, with York time $g_{ij}
\Pi^{ij} = 0 $.

By combining the above conditions, we obtain a new action without time
derivatives, demonstrating that the propagation of the fields
$\alpha$, $\beta$ , $\phi$ can be made instantaneous, as it appears
from the equations of motion of Ref. \cite{a11}:

\begin{eqnarray}
  \nabla^2 \phi + \frac{e^{2\phi}}{\alpha^2} \partial_z {\overline \beta}
\partial_{\overline z} \beta & = & \nabla^2 \phi + N {\overline N}
e^{-2\phi} = - |g| e^{-2{\phi}} T^{00}, \nonumber \\ 
  \partial_{\overline z} \left( \frac{e^{2\phi}}{\alpha} \partial_z
{\overline \beta} \right) & = & \partial_{\overline z} N = 
- \half\alpha^{-1} |g|( T^{0z} - \beta T^{00} ), \nonumber \\ 
  \nabla^2 \alpha - 2 \frac{e^{2\phi}}{\alpha} \partial_z
{\overline \beta} \partial_{\overline z} \beta & = &  \alpha^{-1} |g| 
( T^{z{\overline z}} - \beta T^{0{\overline z}} - {\overline \beta} T^{0z} + 
\beta {\overline \beta} T^{00}).\end{eqnarray}

We understand from Eq. (2.26) that the sources of the meromorphic $N$ 
function are given by a combination of $\delta$-functions. For two
particles, this leads to the solution of [BCV],
\beq N = \frac{C {(\xi_{21})}^{-1 -\frac{\cal M}{2\pi}} }{\zeta ( 1 - \zeta )}
\eeq
which shows simple poles at $z = \xi_1$ and $z= \xi_2$, and no pole at $\zeta = 
\infty$. The $N$ function so determined provides also a feedback in
the first of Eqs. (2.26),
in which it modifies the sources of the ``Liouville field'' 
$e^{2{\widetilde\phi}} = e^{2\phi} / {|N|}^2 $, which is determined by the 
mapping function $f$ in Eq. (2.9).

The $t$-dependence of the trajectories is now provided by the covariant
conservation of the energy-momentum tensor, which in turn implies the
geodesic equations

\beq \frac{d^2 \xi_i^\mu}{ds^2_i} + {(\Gamma^\mu_{\alpha\beta})}_i 
\ \frac{d\xi^\alpha_i}{ds_i} \ \frac{d\xi^\beta_i}{ds_i} = 0 \ \ \ i = 1,
..., N \eeq

Remarkably, we can completely solve these geodesic equations in the
first-order formalism, by measuring the distance between two particles
in the $X^a$ coordinates:

\begin{eqnarray}
 X^a_2 (t) - X^a_1 (t) & = & B^a_2 - B^a_1 + V^a_2 T_2 - V^a_1 T_1 =
\nonumber \\ 
& = &\ \int^{\xi_2}_{\xi_1} \ dz \ N \ W^a ( z, t) + 
\int^{{\overline\xi}_2}_{{\overline\xi}_1} \ d {\overline z} 
{\overline N} \ {\widetilde W}^a ( {\overline z}, t). \end{eqnarray}

The explicit solution is obtained by inverting these relations to
obtain the trajectories  $\xi_i(t)$ as functions of the
constants of motion $B_i^a$, $V^a_i$.
 
{\bf 2.3 Spinning particle metric }

The metric of a spinning particle at rest \cite{a6} is related to the large
distance behaviour of the one of two moving particles which carry
orbital angular momentum $J$. In the latter case,from the two-body solution 
\cite{a12}, we find that the Minkowskian time $T$ at large distance from
the sources changes by

\beq \Delta T = - 8 \pi G J = - J  \ \ \ \ \ \ \ ( 8 \pi G = 1 ) \eeq
when the angular variable $\theta$ changes by $2\pi$, so that we must identify
times which differ by $8 \pi G J $ to preserve single-valuedness of the $X^a$ 
mapping.

Analogously, it was realized long ago \cite{a6} that $T$ has a shift proportional to 
the spin $s$ , when turning around a spinning source, while the spatial
component $Z$ rotates by the deficit angle $m$. By the transformation

\beq T = t - \frac{s}{2\pi} \theta , \ \ \ 
\ \ \ Z = \frac{w^{1-\mu}}{1-\mu} , \ \ \ \ \mu = \frac{m}{2\pi} , \eeq

the Minkowskian metric becomes the one of Ref . \cite{a6}:

\beq ds^2 = {\left( dt - \frac{s}{2\pi} d \theta \right)}^2 - {|w|}^{-2\mu} dw
d{\overline w} \eeq
which, however, is not conformal ( $ g_{zz} \neq 0 $ ). 

Hence, in order to switch to a conformal type gauge, we must modify the $Z$ 
mapping of Eq. (2.31) by allowing another term, dependent on $\overline z$ ,
which preserves the polydromy properties of the first term. If we set 

\beq Z = \frac{1}{1-\mu} \left( z^{1-\mu} + A^2 \ {\overline z}^{\mu -
1} \right) + B \eeq
(where $B$ is an additional arbitrary constant), we can get a
cancellation of the $g_{zz}$ term in the metric \cite{a13}:

\beq ds^2 = {\left( dt - \frac{s}{4\pi} \left( \frac{dz}{iz} -
\frac{d{\overline z}}{i {\overline z}} \right) \right)}^2 - 
{| z^{-\mu} dz - A^2 {\overline z}^{\mu-2} d{\overline z} |}^2 \eeq
by choosing $A \ = \ - s / 4\pi$. 

The one-particle solution now looks like:

\beq ds^2 = dt^2 + \frac{s}{2\pi} ( dt \ i \ \frac{dz}{z} + h.c. ) -
e^{2\phi} dz d {\overline z} \eeq
where the conformal factor $e^{2\phi}$ is given by:

\beq e^{2\phi} = {|z|}^{-2\mu} {( 1 - A^2 {|z|}^{2(\mu-1)} )}^2 =
{|\frac{N}{
f'}|}^2
{( 1 - {|f|}^2)}^2 \eeq
In our general solution of Eqs. (2.8)-(2.10) this expression implies the
following choice for the analytic functions $N(z)$, $f(z)$:

\begin{eqnarray}
N(z) & = & - \frac{i A ( \mu - 1) }{z^2} \ , \ \ \  \mu \equiv \frac{m}{2\pi} 
\nonumber \\  f(z) & = & - i A z^{\mu -1} 
\end{eqnarray}

Let us first note that $f$ has a singular behaviour for $z\rightarrow 0$, 
compared to the one of Eq. (2.11), and takes therefore large values for $z$ small.
This implies that in $z$ coordinates, there is a ``horizon'' surrounding the
spinning particle which corresponds to $|f| = 1$, and thus to a vanishing 
determinant $\sqrt{|g|} \simeq \alpha e^{2\phi} = 0 $ in
Eq. (2.36). It is therefore given by the circle

\beq {|z|}^2 = r^2_0 = A^{\displaystyle{\frac{2}{1-\mu}}}, \eeq
or, in $Z$ coordinates, by the circle

\beq Z_0 = \frac{2A}{1-\mu} e^{i\theta (1-\mu)} \Rightarrow |Z_0| =
R_0 = \frac{s}{2\pi (1-\mu)}.\eeq

The meaning of this ``horizon'' is that values of $Z$ with $|Z| < R_0$ cannot
be obtained from the parametrization (2.33) for any values of
$z$. This in 
turn is related to the fact that, due to the symmetry $ z \rightarrow
A^{\frac{2}{1-\mu}} \ / \ {\overline z} $, the inverse of the mapping
(2.33) is not single-valued , and in fact we shall choose the
determination of $z$ such that $z \simeq Z^{ \ \frac{1}{1-\mu}}$ for $Z 
\rightarrow \infty$.

The fact that our gauge is unable to describe the internal region $|Z|
< R_0$ is related to the existence, in that region, of CTC's
\cite{a14}, which do
not allow a global time choice. Indeed, closed time-like curves can be
built when the negative time-jump $\Delta T = - s$ cannot be
compensated by the time occurring to a light signal to circle the
particle at distance $R$, which is given by $T_{travel} = 2\pi ( 1 - \mu ) R$,
thus implying

\beq  R < \frac{s}{2\pi (1-\mu)} = R_0 , \eeq
i.e., the same critical radius as in Eq. (2.39). For this reason we
shall call the sort of horizon just found a ``CTC horizon''.

Secondly, we note that the ratio $ N /{f'}$ is similar to the
spinless case, but the behaviour of $N(z)$ and $f'(z)$ separately is
more singular, i.e. 

\beq N \simeq \frac{\sigma}{z^2} , \ \ \ f' \simeq z^{\mu-2}. \eeq 

This is to be expected because the source for $N(z)$ in the equation
of motion (2.26) is a more singular distribution, i.e.

\beq \partial_{\overline z} N(z) \propto \frac{s}{2\pi} \delta' (r)
\eeq
in order to allow for a localized angular momentum.  The same
$z^{-2}$ behaviour for $N(z)$ can be found in the large
distance limit of the two-body problem.

This more singular behaviour of the metric is an obstacle to define
the $X^a$ coordinates in the vicinity of the particle at
rest. Nevertheless the particle site can be unambiguously obtained by looking
at the center of rotation of the DJH matching conditions \cite{a2} arising
from (2.33), when turning around $z=0$ in the region outside the CTC
horizon, i.e.,

\beq  Z - B \rightarrow e^{-2i \pi \mu } ( Z - B ). \eeq

In the following, we shall use this procedure in order to identify the
value of the $Z$ coordinate at the particle site (Cfr. Appendix A ).

\section{Spinning particles at rest}

The function $N(z,t)$ plays an important role in the following
discussion because
its polar structure determines the time shift around each particle,
and from this we can get information about the apparent singularities
which appear in the spinning case.

Let us recall the form (2.27) of $N(z)$ for the two-body problem in
the spinless case:

\beq N(z,t) = - \frac{R(\xi(t))}{(z-\xi_1)(z-\xi_2)} = \frac{R(\xi)}{\xi^2}
\frac{1}{\zeta (1- \zeta)} \eeq
where $\xi \equiv \xi_{21} = \xi_2 - \xi_1 $ is the interparticle
coordinate, and $R(\xi)$ was determined \cite{a12} to be

\beq R(\xi) = C \xi(t)^{1-\frac{\cal M}{2\pi}} \eeq

The imaginary part of this coefficient is related to the asymptotic time shift by

\beq \Delta T \simeq - R \int \frac{dz}{z^2} \frac{f}{f'} + (h.c.) =  
- \frac{4\pi}{1 - \frac{\cal M}{2\pi}} Im R = - J \eeq

where $\cal M$ is the total mass of Eq. (2.12).

Therefore, by Eq. (2.30) $R$ is determined in terms of the total
angular momentum of the system, which for the spinless case is purely
orbital ($J = L$), and given by \cite{a12}

\beq 2 \gamma_{12} |V_2 - V_1| \ B_{21} \ \frac{\sin\pi\mu_1 \sin\pi\mu_2}{\sin
\frac{\cal M}{2} } = L = \frac{4\pi}{1-\frac{\cal M}{2\pi}} \ Im (R) \eeq

In the spinning case, we can assume that at large distances $N(z,t)$ has the
same $z^{-2}$ behaviour as in the spinless case. However, around each particle
$N$ should have double poles, as found in Eqs. (2.37) and (2.41). We take
therefore the ansatz

\beq N(z,t) = \frac{R(\xi )}{\xi^2} \left( \frac{1 - \sigma_1 - \sigma_2}{
\zeta ( 1 - \zeta )} - \frac{\sigma_1}{\zeta^2} - \frac{\sigma_2}{{(\zeta -1 
)}^2} \right) , \eeq

where $R\sigma_i$ are the double pole residues.

As a consequence of the double poles, logarithmic contributions to $T$ appear
around each particle, which give rise to a time shift proportional to each
spin $s_i$:

\beq \Delta T_i = \oint_{C_i} \ dz \ \frac{Nf}{f'} + (h.c) = - 4 \pi \frac{
Im ( R \sigma_i)}{1 - \mu_i} = - s_i , \ \ \ \ \ \  
\mu_i \equiv \frac{m_i}{2\pi} \eeq
where we have used the power behaviour $f|_i \simeq z^{\mu_i-1}$.
Eqs. (3.6) and (3.3) determine the values of the $\sigma_i$'s in terms of spins
and angular momentum:

\beq \sigma_i = \frac{(1- \mu_i)}{(1 - \frac{\cal M}{2\pi}) } 
\frac{s_i}{J} \eeq

Furthermore, eq. (3.5) can be rewritten so as to show the presence of two
zeros of $N$, at $\zeta = \eta_1$, $\zeta =\eta_2$, i.e.,

\beq N(z,t) = - \frac{R(\xi)}{\xi^2} \frac{(\zeta-\eta_1)(\zeta-\eta_2)}{
\zeta^2 {(\zeta-1)}^2 } \eeq

with

\beq \eta_{1,2} = \half \left( 1 +\sigma_1 -\sigma_2 \pm \sqrt{ 1 - 2 ( 
\sigma_1 + \sigma_2 ) + {(\sigma_1 - \sigma_2)}^2 } \right) \eeq

Such zeros turn out to be the apparent singularities of our system. In fact,
in order to avoid zeros of the metric determinant we have to cancel them by 
having $f'$ to vanish at $\zeta = \eta_i$ too. Therefore, even if $f$
is analytic around $\zeta = \eta_i$, its Schwarzian derivative has
extra double poles with differences of indices ${\tilde \mu} = 2$,
corresponding to the general parametrization

\begin{eqnarray}
 \{ f, z \} & = & \half \frac{ \mu_1 ( 2 - \mu_1 )}{\zeta^2} + \half 
\frac{ \mu_2 ( 2 - \mu_2 )}{{(\zeta - 1)}^2} - \frac{3}{2} \frac{1}{ {(\zeta -
\eta_1)}^2} - \frac{3}{2} \frac{1}{ {(\zeta - \eta_2)}^2} + \nonumber \\
& + & \frac{\beta_1}{\zeta} + \frac{\beta_2}{\zeta -1} + 
\frac{\gamma_1}{\zeta-\eta_1} + \frac{\gamma_2}{\zeta-\eta_2}  \end{eqnarray}
with the ``accessory parameters'' $\beta's$ and $\gamma's$ so far undetermined.

{\bf 3.1 The two-body case }

There is no general known solution to the Fuchsian problem of Eq. (3.10)
, since it contains five singularities. However in a few limiting 
cases a particular solution can be obtained. For example in the static
two-body case under consideration, the
form of $f'$ is determined by its zeros, and by the known behaviour
around $\zeta = 0$ and $\zeta = 1 $, as follows :

\beq f' = - \frac{K}{\xi} (\zeta-\eta_1)(\zeta-\eta_2)
\zeta^{\mu_1-2} {(1-\zeta)}^{\mu_2-2}. \eeq
Furthermore, it should be integrable to a function $f$ with static 
monodromy matrix, and behaviour $f \simeq \zeta^{\mu_1 + \mu_2 - 1}$
at $\zeta = \infty$, of the form:

\beq f = \frac{K}{\mu_1+\mu_2-1} \zeta^{\mu_1-1} {(1-\zeta)}^{\mu_2-1} 
(\zeta-\tau). \eeq
The consistency of Eqs. (3.11) and (3.12) gives a constraint on the
possible values of $\eta_1$, $\eta_2$:

\beq \eta_1 \eta_2 = \sigma_1 = \frac{(1-\mu_1) \tau}{1-\mu_1-\mu_2} \
, \ \ \ \ ( 1 - \eta_1 ) ( 1 - \eta_2 ) = \sigma_2 =
\frac{(1-\mu_2)(1-\tau)}{1-\mu_1-\mu_2} \eeq
which, by Eq. (3.7), is satisfied if the total angular momentum is 
simply the sum of the two spins $s_i$, i.e. $ J = s_1 + s_2 =
S$, or

\beq \frac{\sigma_1}{1-\mu_1} + \frac{\sigma_2}{1-\mu_2} = \frac{1}{
1-\mu_1-\mu_2} , \eeq
as expected in the static case. This condition also determines the
value of 

\beq R = i \frac{S}{4\pi} \left( 1 - \mu_1 - \mu_2 \right) \eeq

In order to determine the constant $K$ in Eq. (3.12) we must use the
analog of Eq. (2.29) which defines the Minkowskian interparticle distance

\begin{eqnarray}
& B_{21} & =  Z_2 - Z_1  = \int^2_1 \ dz \ \frac{N}{f'} + 
\int^2_1 \ d{\overline z} \
\frac{{\overline N}{\overline f}^2}{\overline f'} = 
\left( \frac{R}{K} \int^1_0 \ d\zeta \ \zeta^{-\mu_1} {( 1 - \zeta
)}^{-\mu_2} + \right. \nonumber \\
& + & \left. \frac{{\overline R} {\overline K}}{{(\mu_1 + \mu_2 -1)}^2}
 \int^1_0  \ d\zeta \
\zeta^{\mu_1 -2} {(1-\zeta)}^{\mu_2 -2} {(\zeta-\tau)}^2\right)
\end{eqnarray}

We can see that the second integral is not well defined in the
physical range

\beq 0 < \mu_i < 1 \ \ , \ \ \ \ \ \ 0 < \mu_1 + \mu_2 < 1 \eeq
for which, as shown in I, there are no CTC's at large distances. This
fact reflects the existence of $CTC$ horizons close to the particles,
(cfr. Sec. (2.3)), in which the mapping to Minkowskian coordinates is
not well defined.

The rigorous way of overcoming this problem is to solve for the DJH
matching conditions of type (2.43) outside the $CTC$
horizons, thus defining $B_{21}$
as the relevant translational parameter, as explained in Appendix A.

Here we just notice that the integral in question can be defined by
analytic continuation from the region $ 2 > \mu_i > 1$ to the region
(3.17), so as to yield, by Eqs. (3.15) purely imaginary values for $K$
and $R$, with

\begin{eqnarray}
& \ & \frac{1}{|K|}  B ( 1-\mu_1, 1-\mu_2 ) - \frac{|K|
B(\mu_1 , \mu_2 )}{{(\mu_1+\mu_2-1)}^2}
 \left( 1 - \tau \frac{1-\mu_1-\mu_2}{1-\mu_1} -
(1-\tau) \frac{1-\mu_1-\mu_2}{1-\mu_2} + \right. \nonumber \\
& + &  \tau ( 1 - \tau ) \left.
\frac{(1-\mu_1-\mu_2)(2-\mu_1-\mu_2)}{(1-\mu_1)(1-\mu_2)} 
\right) =  \frac{4\pi \ B_{21}}{(1-\mu_1 -\mu_2) S} 
\end{eqnarray}
where $S = J = s_1 + s_2$ denotes the total spin. In this equation the smaller
branch of $K$ should be chosen for the solution to satisfy $|f|<1$
closer to the particles.

In particular, if $S / B_{12} \ll 1$, the acceptable branch of
the normalization $K$ becomes small and of the same order. In general,
however $K$ is not a small parameter, and thus $f$ is not small,
unlike the spinless case in which $f$ is of the order of
$ L / B_{12}$ and is thus infinitesimal in the static limit.

Whatever the value of $K$, for $z$ sufficiently close to the
particles, the critical value $|f|= 1$ is reached, because of the
singular behaviour of Eq. (3.12) for $\mu_i < 1$. Therefore we have two
horizons encircling each particle, which may degenerate in one
encircling both for sufficiently large values of $S/B_{12}$.

Having found an explicit solution for $f$, its Schwarzian derivative
is easily computed from Eq. (2.14). By using the notation 

\begin{eqnarray} {\tilde \mu}_i & = & \mu_i - 1 \ \ \ ( i = 1,2 ) \ , \ \
\ \ {\tilde \mu_3} = {\tilde \mu_4} = 2 \nonumber \\
\eta_1 & = & \zeta_3 , \ \ \eta_2 = \zeta_4 , \ \ \ \ \ \  \gamma_1 =
\beta_3 \ , \ \ \ \gamma_2 = \beta_4 \end{eqnarray}
we find that the residues $\beta_i$ at the single poles of Eq. (3.10)
( called the accessory parameters) take the quasi-static form of BCV, i.e.
 
\beq \beta_i = - ( 1 - {\tilde \mu}_i ) \sum_{j \neq i} \frac{(1 -
{\tilde \mu}_j ) }{\zeta_i - \zeta_j} \ \ , \ \ \ \ \ \ ( i, j = 1, ...., 4 ). \eeq

{\bf 3.2 The static metric}

From the two-body static solution for $f$ and $N$ we can build the
static vierbein of Eqs. (2.5) and (2.15) which has the components:

\begin{eqnarray} 
& E_0^a & =  ( 1 , 0 , 0 ) \nonumber \\
& E^a_z & =  N W^a = \frac{R }{\xi (\mu_1+\mu_2-1)}  \left(
\frac{\zeta-\tau}{\zeta (1-\zeta)} , \frac{\mu_1+\mu_2-1}{K_0}
\zeta^{-\mu_1}{(1-\zeta)}^{-\mu_2} , \right. \nonumber \\
&& \frac{K_0}{(\mu_1+\mu_2-1)} \left. 
\zeta^{\mu_1-2} {(1-\zeta)}^{\mu_2-2} {(\zeta-\tau)}^2 \right) 
\end{eqnarray}

The corresponding metric has the form

\begin{eqnarray}
g_{00} &=& 1 \nonumber \\
g_{0z} &=& \frac{Nf}{f'} = \frac{R}{\xi (\mu_1+\mu_2-1)}
\frac{\zeta-\tau}{\zeta(1-\zeta)} \nonumber \\
-2 g_{z\overline{z}} & = & e^{2\phi} = \frac{R^2}{\xi^2 |K_0|^2} 
{(\zeta{\overline\zeta})}^{-\mu_1}
{((1-\zeta)(1-{\overline\zeta}))}^{-\mu_2} \cdot \nonumber \\
&\cdot&
{\left[ 1 - \frac{K^2_0 {(\zeta\overline\zeta)}^{\mu_1-1}}{{(\mu_1+\mu_2-1)}^2}
{((1-\zeta)(1-\overline\zeta))}^{\mu_2-1}
(\zeta-\tau)(\overline\zeta-\overline\tau) \right]}^2 
\end{eqnarray}

and is degenerate whenever $e^{2\phi} = 0$, revealing explicitly the
presence of a singularity line on which the determinant of the metric
is vanishing.  

Let us remark that the zeros of $f'$, the apparent singularities, are 
geometrically saddle points for the modulus ${|f|}^2$, which instead diverges
on the particle sites. It is easy to realize that in the range $ S/
B_{12} \ll 1$, where $K_0 \simeq S / B_{12}$, the curve $|f|=1$ defines two
distinct horizons, one for each particle. 

In the complementary range $ S \ge B_{12} $, with $K_0$ satisfying (3.18)
in its generality , the curve $|f|=1$ defines a line surrounding both 
particles.  

As a conseguence, we can distinguish the two particles and set up the 
scattering problem only in the case where $ S/B_{12}$ is at most 
of order O(1). This restriction is physically motivated by the
presence of closed timelike-curves, which make impossible to reduce
the impact parameter without entering in causality problems.

{\bf 3.3 The N-body static case }

In the general static case, with $N$ bodies, we can also provide a
solution for the mapping function by algebraic methods, following the
pattern described above.

Firstly, the meromorphic $N$ function, having simple and double poles
at $\zeta = \zeta_i$ can be parametrized as
 
\beq N = R \left(  - \sum^N_{i=1} \frac{\sigma_i}{{(\zeta-\zeta_i)}^2}
+ \sum^N_{i=1} \frac{\nu_i}{\zeta-\zeta_i} \right) = R \frac{
\prod^{2N-2}_{\xi=1} ( \zeta-\eta_i ) }{ \prod^N_{i=1}
{(\zeta-\zeta_i)}^2} \eeq
with the following conditions

\begin{eqnarray}
 \sum_i \nu_i & = & 0 \ , \ \ \ \ ( N \sim - R z^{-2} \ {\rm for } \  z
\rightarrow \infty ) \nonumber \\ 
\sum_i \sigma_i & - & \sum_i \zeta_i \nu_i = 1 \ , \ \ \ \ 
( \ {\rm normalization \ of } \ R \ ). \end{eqnarray}

Therefore, there are $2N-2$ zeros ( or apparent singularities ) given
in terms of the $\sigma$'s and of $N-2$ $\nu$-type
parameters. Furthermore, the $\sigma$'s are given by the time shifts
in terms of $s_i / J$ as in Eq. (3.7).

Secondly, $f'$ shows the same $2N-2$ zeros at $\zeta = \eta_j$ in the form
 
\beq \frac{df}{dz} = \frac{1}{\xi} f'(\zeta) = \frac{K}{\xi}
\ \prod^{N}_{i=1} \ {( \zeta - \zeta_i )}^{\mu_i -2}
\prod^{2N-2}_{j=1} \ {(\zeta - \eta_j)} \eeq

while the mapping function, having static monodromy and behaviour
$\zeta^{\sum_i \mu_i - 1}$ at $\zeta = \infty$, has only $(N-1)$ zeros
with the form

\beq f = \frac{K}{\sum_i \mu_i - 1} \ \prod^N_{i=1} {(\zeta-
\zeta_i)}^{\mu_i-1} \ \prod^{N-1}_{k=1} (\zeta - \tau_k) . \eeq

At this point, the integrability condition, that (3.25) is just the
$z$-derivative of (3.26) provides $2N-2$ conditions for a total of
$N-2 + N-1 = 2N-3$ parameters. Therefore, all the $\eta$ parameters
are determined as function of the $\sigma$'s, and there is one extra
condition among the $\sigma$'s, namely that

\beq \sum^n_{i=1} \ \frac{\sigma_i}{1-\mu_i} = \frac{1}{1- \sum_i
\mu_i} \eeq
which is verified, by Eq. (3.7) because $\sum_i s_i = S = J$ in the
static case. 
 
Finally, the normalization $K$ and the $N-2$ ``shape parameters''
$\zeta_j = \frac{\xi_{j1}}{\xi_{21}}$ are determined from the $N-1$ 
``equations of motion'' 

\beq B_j - B_1 = \int^j_1 \ \frac{N}{f'} \ dz \ + \ \int^j_1 \
\frac{{\overline N}{\overline f}^2}{\overline f'} , \eeq
similarly to the two-body case.

We conclude that the static $N$-body case with spin provides a
solvable example of non-vanishing mapping function with static
monodromies and total mass ${\cal M} = \sum_{i=1}^N m_i$, having a
Schwarzian with a total of $3N-1$ singularities.

\section{Spinning particles in slow motion}

For two moving spinning particles, the fuchsian Riemann-Hilbert
problem for the mapping function is in principle well defined by
Eqs. (2.13), (2.14) and (3.10). Indeed, the location of the apparent
singularities is fixed in general by Eqs. (3.5), (3.8) and (3.9) and
it is possible to see that all accessory parameters in the Schwarzian
of Eq. (3.10) are also fixed in terms of the invariant mass ${\cal M}$
of Eq. (2.12) and of the spins.

The fact that the potential of the Fuchsian problem is determined
follows from some general conditions that the accessory parameters
should satisfy, which were described in I, and are the following.

Firstly, the point at $\zeta = \infty$ is regular, with difference of
exponents given by $\mu_\infty = \frac{\cal M}{2\pi} - 1$. This yields
two conditions:

\begin{eqnarray} 
& \ & \sum^2_{i=1}  \ ( \beta_i + \gamma_i )  \ = \ 0 , \nonumber \\
& \ & \sum^2_{i=1}  \ \mu_i ( 2 - \mu_i ) - 6 + 2 \sum^2_{i=1} \ ( \beta_i
\zeta_i + \gamma_i \eta_i ) = 1 - \mu^2_\infty . 
\end{eqnarray}

Secondly, there is no logarithmic behaviour \cite{a16} of the solutions $y_i$
at the apparent singularities $\eta_j$. This yields two more
conditions:

\begin{eqnarray}
- \gamma^2_1 \ = \ \sum^2_{j=1} \frac{\mu_j (2-\mu_j)}{{(\eta_1 - \zeta_j)}^2} 
- \frac{3}{{(\eta_1 - \eta_2)}^2} + \sum_j \frac{2 \beta_j}{\eta_1 -
\zeta_j} + \frac{2\gamma_2}{\eta_1-\eta_2} , \nonumber \\
- \gamma^2_2 \ = \ \sum^2_{j=1} \frac{\mu_j (2-\mu_j)}{{(\eta_2 - \zeta_j)}^2} 
- \frac{3}{{(\eta_1 - \eta_2)}^2} + \sum_j \frac{2 \beta_j}{\eta_2 -
\zeta_j} + \frac{2\gamma_1}{\eta_2-\eta_1}.
\end{eqnarray}

The four algebraic ( non linear ) Eqs. (4.1) - (4.2) determine
$\beta_1$, $\beta_2$ and $\gamma_1$, $\gamma_2$ in terms of ${\cal M}$
and of the $\sigma$'s , similarly to what happens in the ( much
simpler ) spinless case. However, since no general solution to this
Fuchsian problem with $5$ singularities is known, one should resort to
approximation methods in order to provide the mapping function
explicitly.

The idea is to expand $f$ in the (small) Minkowskian velocities $V_i
<< 1$, around the static solution, which is exactly known. One should,
however, distinguish two cases, according to whether $ S / B_{12} \ll
1$ is of the same order as the $V$'s, or instead $ S / B_{12} = O(1)
$, where $B_{12}= B_1 - B_2$ is defined as the relative impact parameter of the
Minkowskian trajectories

\beq Z_1 = B_1 + V_1 T_1 \ \ \ \ \ \ Z_2 = B_2 + V_2 T_2 . \eeq

In the first case, that we call ``peripheral'',
the expansion we are considering is effectively in
both the $V_i$'s and $f$ itself, at least in a region sufficiently far
away from the horizons, which do not overlap, as noticed before. This
is a situation of peripheral scattering with respect to the scale
provided by the horizon, and will be treated in the following to first
(quasi-static) and second ( non-relativistic ) order.

The second case, that we call ``central'', 
($S / B_{12} = O(1)$) has non-linear features even to
first order in the velocities and the horizons may overlap. We shall
only treat the quasi-static case.

{\bf 4.1 Peripheral quasi-static case ( $S  \ll B_{12}$ )}

Since in this case both $f$ and $V_i$ can be considered as small, at
first order we have just to look for a mapping function which solves
the linearized monodromies of Eq. (2.7) around the particles (i = 1,2)

\beq {\tilde f}_i = \frac{a_i}{a^{*}_i} f (\zeta) +
\frac{b_i}{a^{*}_i} \ \ \  \ \ ( a_i = e^{i \pi \mu_i} \ , \ \ b_i = -
i {\overline V}_i \sin \pi \mu_i ) \eeq

and, furthermore, has the two apparent singularities in Eq. (3.9) and
the behaviour in Eq. (2.41) at $\zeta =0$ and $\zeta = 1$. Since, by
Eq. (4.4), $f'$ has static monodromies, we can set

\beq f' = K \zeta^{\mu_1-2} {(1-\zeta)}^{\mu_2-2} (\zeta-\eta_1)
(\zeta-\eta_2), \eeq

as in Eq. (3.11).

However, in the moving case, $f'$ is not integrable and $f$ has the
quasi-static monodromy (4.4), which contains a translational part. We
then set

\beq f(\zeta) = f(i) + \int^{\zeta}_i \ dt \ f' (t) \eeq
where the ith-integral is understood as analytic continuation from
$\mu_i > 1$, as explained in Appendix A.

Since the ith-integral has purely static monodromy around the i-th particle
we obtain from
Eq. (4.6)

\beq ( {\tilde f}(\zeta) - f(i) ) = e^{2 i \pi \mu_i} ( f (\zeta) - f(i)
) \eeq

and thus Eq. (4.4) is satisfied if $f(i) = - \frac{\overline V_i}{2}$,
thus yielding the condition

\begin{eqnarray} 
\frac{{\overline V}_2 - {\overline V}_1}{2} & = & K \int^1_0 \ dt \
t^{\mu_1 -2} {(1-t)}^{\mu_2 -2} (t-\eta_1) (t-\eta_2) =  \nonumber \\
& = & K B(\mu_1, \mu_2 ) \left( 1 - \sigma_1
\frac{1-\mu_1}{1-\mu_1-\mu_2} - \sigma_2 \frac{1-\mu_2}{1-\mu_1-\mu_2} \right)
\end{eqnarray}

By then using Eq. (3.7), we determine the normalization

\beq K = \frac{{\overline V}_{21}}{2} B^{-1}(\mu_1, \mu_2) {\left[ 1 -
\frac{s_1+s_2}{J} \right]}^{-1} = B^{-1}(\mu_1, \mu_2) \frac{J}{L}
\frac{{\overline V}_{21}}{2} \eeq

where $L$ is the orbital angular momentum of the system, given in
Eq. (3.4).

Eqs. (4.5), (4.6) and (4.9) yield the ( peripheral ) quasi-static solution with
spin. In particular, Eq. (4.9) shows the existence of two regimes,
according to whether the spin is small or large with respect to the
orbital angular momentum $L$.

If the spin $S$ is small ( $S \ll L$ ), so are the $\sigma$'s, the
apparent singularities of Eq. (3.9) become degenerate with the particles

\beq \eta_1 \simeq \sigma_1 , \ \ \ \eta_2 \simeq 1 - \sigma_2 \eeq
and the normalization $K \sim {\overline V}_{21}$ is vanishingly small
in the static limit, thus recovering the spinless quasi-static limit
of I.

If instead $B_{21} \gg S \gg L$ the parameters $\sigma$ and $\eta$ are
of order unity, and , by Eq. (3.4), the normalization becomes

\beq K \simeq B^{-1}(\mu_1,\mu_2) \frac{S}{2L} {\overline V}_{21} = 
\frac{S}{4\pi} \frac{(1-\mu_1-\mu_2) B(1-\mu_1,1-\mu_2)}{B_{21}} , \eeq
in agreement with the static case relation (3.18) for $S \ll
B_{21}$. In the latter case the mapping function becomes nontrivial in
the static limit, as discussed in the previous section.

In the general case, for any $S/L$ of order unity, the mapping
function $f$ is of first order in the small parameters and its
Schwarzian derivative does not change with respect to the static case,
except for the actual values of the $\sigma$'s and $\eta$'s, so that
the accessory parameters are provided by Eq. (3.20). The first
non-trivial change of $\{ f , \zeta \}$ is at second order in the small
parameters, as we shall see ( Secs. 4.2 and 4.3 ).

{\bf 4.2 Central quasi-static case ( $S / B_{12} \simeq O(1)$) }

In this case we have to expand the monodromies in Eq. (2.7) in the
$b_i$ parameters only, thus keeping possible non linear terms in
$f$. By expanding around the quasi-static solution $f^{(0)}$ of Eq. (4.5)
we can write

\beq f = f^{(0)} + \delta f + .... \eeq

where, around the generic particle,

\beq {\tilde f} = \frac{a_i}{a^{*}_i} f + \frac{b_i}{a^{*}_i} -
\frac{a_i b^{*}_i}{{a*}^2_i} {(f^{(0)}_s)}^2 , \ \ \ \ \ \ \ i = 1,2 \eeq
and we have introduced in the last term the static limit $f^{(0)}_s$
of Eq. (3.12). These monodromy conditions, unlike the ones in Eq. (4.4), are
nonlinear. However it is not difficult to check that they linearize for the
function

\beq h = \frac{1}{f^{(0)}_s}\frac{f'}{f'^{(0)}} \ \  , \eeq
which satisfies the first-order monodromy conditions

\beq {\tilde h} = e^{-2 i \pi \mu_i} h - V_i ( 1 - e^{- 2 i \pi \mu_i}
) . \eeq

Furthermore, from the boundary conditions for $f$, we derive the
following ones for $h$

\beq
 h \simeq \left\{ \begin{array}{cc} - V_i + O ( {( \zeta -
\zeta_i)}^{1-\mu_i} ) \ \ , & ( \zeta \simeq \zeta_i, \ i = 1, 2 ) , \\
\zeta^{1-\mu_1-\mu_2} \ \ , & ( \zeta \rightarrow \infty ) \\
{( \zeta - \tau )}^{-1} \ \ , & ( \zeta \rightarrow \tau ) 
\end{array} \right. \eeq

The solution for $h$ can be found from the ansatz

\begin{eqnarray} 
h & = &  - V_1 + ( V_1 - V_2 ) I_0 ( \zeta ) + A \frac{
\zeta^{1-\mu_1} {(1-\zeta)}^{1-\mu_2}}{K_0 ( \zeta - \tau )} \nonumber \\
I_0 & = & \frac{1}{B(1-\mu_1, 1- \mu_2)} \int^{\zeta}_0 \ dt \ t^{-\mu_1}
\ {(1-t)}^{-\mu_2} \end{eqnarray}
by noticing that the first two terms automatically satisfy the
boundary conditions at $\zeta = 0, 1, \infty$. The last term is
proportional to ${(f^{(0)}_s)}^{-1}$, contains the pole at $\zeta = \tau$,
and the constant $A$ is determined so as to satisfy the translational
part of the monodromy (4.13). Similarly to Eq. (4.8) we get the equation
for $A$

\beq \frac{{\overline V}_2 - {\overline V}_1}{2} = A \int^1_0 {f'}^{(0)}
(\zeta) + \int^1_0 d\zeta {f'}^{(0)}_s (\zeta) [ - V_1 f^{(0)}_s
(\zeta) + (V_1-V_2) f^{(0)}_s (\zeta) I_0 (\zeta)]  \eeq
where the integrals are understood as analytic continuations from $2 >
\mu_i > 1$, and the non-linear terms represent higher order
contributions in the parameter $S/B_{12}$.

By using the normalization condition (4.8) we then obtain by an
integration by parts and for velocities along the $x$-axis,

\beq A = 1 - \int^1_0 \ d\zeta \ {(f^{(0)}_s)}^2 ( \zeta ) I^{'}_{0}
\eeq
This means that the coefficient of the first-order quasi-static
solution $f^{(0)}$ in Eq. (4.12) is renormalized by higher orders in 
$S/ B_{12}$.
Furthermore from the expression (4.14) of $f'/ f'_{(0)} = h f^{(0)}_s$ we
obtain the correction to the Schwarzian derivative
\begin{eqnarray}
 \{ f , \zeta \}& - &\{ f, \zeta \}^{(0)} \simeq {(h f^{(0)})''}_s -
\frac{{f''}_{(0)}}{{f'}_{(0)}} {( h f^{(0)}_s )}^{'} = \nonumber \\
&=& \frac{K_0 ( V_1 - V_2 )}{(\mu_1+\mu_2-1)B(1-\mu_1,1-\mu_2)} 
\frac{\zeta-\tau}{\zeta(1-\zeta)} \left( \frac{2}{\zeta-\tau} -
\frac{1}{\zeta-\eta_1} - \frac{1}{\zeta-\eta_2} \right) 
\end{eqnarray}
which turns out to be of order $O(V) \cdot O(S / B_{12})$, i.e. 
formally of 1st order in both parameters.

The quasi-static solution for general spin values just obtained is particularly
interesting because it allows to understand how the trajectory
equations (4.3) can make sense, despite the multivaluedness of the
Minkowskian time.

In fact, we can solve for the mapping from regular to Minkowskian coordinates
by expanding around some arbitrary point $\xi_0 \neq \xi_i$ to get,
instead of Eq. (2.10)

\beq X^a = X^a_0 (t) + \int^{\xi}_{\xi_0} \ dz \ N W^a + \int^{\overline\xi}_{\overline
\xi_0} \ d {\overline z} \ {\overline N} \ {\widetilde W}^a . \eeq

We can then explicitly check that, to first order in the velocities,
the combinations $ Z - V_1 T \ ( Z - V_2 T )$ are well defined
at particle $1$ ( particle $2$ ).

For this we need the static time

\beq T = t + \int^{\zeta}_{\zeta_0} \ dz \ \frac{N f_{(0)}}{f'_{(0)}}
+  (c.c) = t - \frac{R}{1-\mu_1-\mu_2} \left( (1-\tau) \log 2(1-\zeta) +
\tau \log 2\zeta \right)  + (h.c.) \eeq
which shows the logarithmic singularities at $z = \xi_i$ and we also
need the $Z$ coordinate up to first order in $V$

\beq Z = Z_0 (t) + \int^{\zeta}_{\zeta_0} dz \ \frac{N}{f'_{(0)}}
\left( 1 - \frac{\delta f'}{f'_{(0)}} \right)
+ \int^{\overline \zeta}_{\overline \zeta_0} \ d {\overline z} \
\frac{\overline N}{\overline f'_0} \left( {\overline f}^2_{(0)} + 2
{\overline f}^{(0)} \delta {\overline f} - \frac{\delta {\overline f'}}{
{\overline f'}_{(0)}} {\overline f}^2_{(0)} \right) 
\eeq

Since, by Eq. (4.14) and (4.17)
 $ \delta f' / {f'_{(0)}} = - V_1 f^{(0)} + O (
{(\zeta - \zeta_i)}^0)$ and $ \delta {\overline f} = - V_1 / 2 + O (
f^2_{(0)} V )$, the $Z$-coordinate also has logarithmic singularities,
which cancel in the combination

\beq \lim_{\xi \rightarrow \xi_1} ( Z - V_1 T ) = B_1 = Z_0 (t) - V_1
t + \int^{\zeta_1}_{\zeta_0} \ dz \ \frac{N}{f'_{(0)}} ( 1 + O (V)) + 
\int^{{\overline \zeta}_1}_{{\overline \zeta}_0} \ d {\overline z} \ 
\frac{{\overline N}
{\overline f}^2_{(0)}}{{\overline f'}_{(0)}} \ ( 1 + O(V)) \eeq
which is thus well defined. From the similar relation from particle
$2$ and using the expression of $R$ in Eq. (3.2)
we obtain the quasi-static equations of motion

\begin{eqnarray}
 i ( B_1 - & B_2) & - ( V_1 - V_2 ) t = \left( \int^{\xi_1}_{\xi_2} \ dz
\frac{N}{f'_{(0)}} + \int^{\overline \xi_1}_{\overline \xi_2} d
{\overline z} \frac{{\overline N }{\overline f}^2_{(0)}}{{\overline f'}_{(0)}}
\right) ( 1 + O(V)) = \nonumber \\
& = & \alpha \ \xi^{1-\frac{{\cal M}}{2 \pi}} + \beta \ {\overline
\xi}^{1-\frac{{\cal M}}{2\pi}} = \nonumber \\
& = & \frac{C}{{K}_0} \xi^{1-\frac{{\cal M}}{2\pi}} B(1-\mu_1, 1-\mu_2)
+ {\overline C} {\overline K}_0 {\overline \xi}^{1-\frac{{\cal M}}{2\pi}}
\frac{B(\mu_1,\mu_2)}{{(\mu_1 + \mu_2 - 1)}^2}
  ( 1 - \tau \frac{1-\mu_1-\mu_2}{1-\mu_1} - \nonumber \\
& - & (1 - \tau ) \frac{1-\mu_1-\mu_2}{1-\mu_2}
+ \tau ( 1 - \tau ) \frac{(1-\mu_1-\mu_2)
(2-\mu_1-\mu_2)}{(1-\mu_1)(1-\mu_2)} )
\end{eqnarray}
where we have assumed the velocities along the $x$ axis, and the
impact parameters along the $y$ axis. The equation (4.25) can be
inverted to give 
\beq C \xi^{1-\frac{{\cal M}}{2\pi}} = \frac{V_{21} t}{\alpha + \beta} +
i \frac{S}{4\pi} (1-\mu_1-\mu_2) \eeq
From Eq. (4.26) we can learn that the spins renormalize the
constants describing the trajectory but not the exponent determining
instead the scattering angle, which is therefore unaffected, and given
by $ \theta = \frac{\cal M}{2} {( 1 - \frac{\cal M}{2\pi} )}^{-1}$ as
in the spinless case. The constant term in the r.h.s. of Eq. (4.26) is expected
to be proportional to $J$, but only the spin part is here determined,
because we have neglected the O(V) terms in Eq. (4.25).
 
{\bf 4.3 The peripheral non relativistic case ( $S \ll B_{12}$ )}

We now expand the projective monodromy transformations of Eq. (2.7) up to
next nontrivial order in both $f$ and $V_i$. By referring to a generic
particle and by defining

\beq f = f^{(1)} + f^{(3)} + ......., \eeq
with similar notation for $a$'s and $b$'s, we obtain

\begin{eqnarray}
{\tilde f}^{(1)} & = & \left. \frac{a}{a*}\right)^{(0)} f^{(1)} + \left.
\frac{b}{a^*}\right)^{(1)} \ , \nonumber \\
{\tilde f}^{(3)} & = & \left. \frac{a}{a*}\right)^{(0)} f^{(3)} - \left.
\frac{a b^*}{a} \right)^{(1)} {(f^{(1)})}^2 + {\left( \frac{a}{a*} -
\frac{|b|^2}{a*^2} \right)}^{(2)} f^{(1)} + \left. \frac{b}{a^{*}} \right)
. \end{eqnarray}

The first equation yields the quasi-static solution described in
Sec. 4.1. The second equation ( which is down by two orders ) is non
linear, but this time it linearizes for the function

\beq h_1 = \frac{1}{f'^{(1)}} \left( \frac{ f'^{(3)}}{ f'^{(1)} }
\right)' \eeq
which satisfies the 1-st order monodromy conditions

\beq \left. {\tilde h}_1 \right)_i = e^{-2 i \pi \mu_i } h_1 - V_i ( 1
- e^{- 2 i \pi \mu_i} ) . \eeq
Furthermore, from the boundary conditions for $f$ we derive the
following ones for $h_1$

\beq h_1 \simeq \left\{ \begin{array}{cc}
 - V_i + O ( {( \zeta - \zeta_i )}^{2-\mu_i} ) , & \zeta \simeq
\zeta_i, \ i = 1,2 \\
\zeta^{1-\mu_1-\mu_2} \ , \ \ \ \zeta \rightarrow \infty \\
{(\zeta- \eta_i )}^{-1} , \ \ \  \zeta \rightarrow \eta_i \end{array}
\right.\eeq
where the values $\left. h_1\right)_i = - V_i$ are needed to realize
the translational part of the monodromy (4.28).

The solution for $h_1$ can be found from the ansatz

\beq h_1 = - V_1 + ( V_1 - V_2 ) \left[ I_0 (\zeta ) + \frac{1}{B(1-\mu_1, 1
-\mu_2)} \frac{\zeta^{1-\mu_1} {(1-\zeta)}^{1-\mu_2} ( A_1 ( \zeta -1
) + A_2 \zeta )}{(\zeta-\eta_1)(\zeta-\eta_2)} \right] \eeq
by noticing that the first two terms automatically satisfy the
translational part of the monodromy (4.28), so that the third one should
have the purely static monodromies $e^{-2 i \pi \mu_i }$, besides the
poles at $ \zeta = \eta_i $. The constants $A_1$ and $A_2$ can then be chosen
so as to satisfy $h + V_i \sim {( \zeta - \zeta_i )}^{2-\mu_i}$. 
Since, by (3.9), $\eta_1 \eta_2 = \sigma_1$,  $ (1-\eta_1) (1-\eta_2)
= \sigma_2$, we find

\beq A_1 = \frac{\sigma_1}{1-\mu_1} = \frac{s_1}{(1-\mu_1-\mu_2) J} \
, \ \ \ \  A_2 = \frac{\sigma_2}{1-\mu_2} = \frac{s_2}{(1-\mu_1-\mu_2)J} \eeq
thus making the spinless limit particularly transparent.

From the form (4.29) of $h_1$, from its definition and from the form of
$f^{(1)}$ in Eqs. (4.5) and (4.6), we then find the result

\begin{eqnarray}
 \frac{f'^{(3)}}{f'^{(1)}} & = &  - V_1 f^{(1)} + \ {\rm const.}
+ \frac{( 1 -\mu_1 - \mu_2 )}{(1-\frac{S}{J})}
 \frac{\delta {\cal M} }{2\pi} \int^{\zeta}_0 dt
\left( - \frac{A_1}{t} + \frac{A_2}{1-t} + \right. \nonumber \\
& + &  t^{\mu_1-2} \left.
{(1-t)}^{\mu_2-2} (t-\eta_1) (t-\eta_2) \int^t_0 d\tau \tau^{-\mu_1}
{(1-\tau)}^{-\mu_2} \right) , \end{eqnarray}
where we have defined the parameter

\beq \delta {\cal M} = {|V_{21}|}^2 \frac{\sin \pi\mu_1 \sin \pi
\mu_2}{\sin \pi ( \mu_1 + \mu_2 )} = {\left[ {\cal M} - ( m_1 + m_2 ) 
\right]}^{(2)} , \eeq
representing the nonrelativistic contribution to the total invariant
mass. In fact, the perturbative large $\zeta$ behaviour of $f'$
provided by Eq. (4.31) is

\beq \frac{f'_{(1)} + f'_{(3)}}{f'_{(1)}} \sim 1 + \frac{\delta {\cal
M}}{2\pi} \log \zeta \sim \zeta^{\frac{\delta {\cal M}}{2\pi}} \eeq
as expected from the behaviour $f' \sim \zeta^{\frac{\cal M}{2\pi} -
2}$ of the full solution.

We further notice that Eq. (4.31) provides a  non-trivial change of the
Schwarzian derivative, similarly to what was noticed in the previous
sections. Since

\beq \{ f , \zeta \} = L'' + \half L'^2 \ , \ \ \ \ \ L \equiv \log
f'^{(1)} + \frac{f'^{(3)}}{f'^{(1)}} + ... \eeq
we have, after some algebra, the non-relativistic correction to the 
Schwarzian (Cfr. App. B)

\begin{eqnarray}  \{ f , \zeta \} & - &  
\{ f , \zeta \}^{(0)} = f'^{(1)} h'_1 +
... = \nonumber \\ & = & \frac{\delta {\cal M}}{2\pi} \frac{1}{1- \frac{S}{J}}
\left[ -
\frac{1}{\zeta (1 - \zeta )} \left( 1 - \mu_1 - \mu_2 - ( 3 - \mu_1 - \mu_2
)\frac{s_1 + s_2}{J} \right) + \right. \nonumber \\
 & + & \left. \left( \frac{s_1}{J} \frac{1}{\zeta} - 
\frac{s_2}{J} \frac{1}{1-\zeta} \right) \left( \frac{1}{\zeta-\eta_1} +
\frac{1}{\zeta-\eta_2} \right) \right] , \end{eqnarray}
where $\{ f, \zeta \}^{(0)}$ denotes the static expression in Eqs. 
(3.10) and (3.20).

The same expression (4.38) could have been obtained by expanding 
Eqs. (4.1)-(4.2) in the parameter $\delta {\cal M}$, around the static
solution of Eq. (3.20) ( Appendix B).

Let us note that the solution here considered contains more
information than simply the terms of order $O(V^2)$ , because the expansion
of the  $1 - \frac{S}{J}$
denominator can give rise in the small velocity limit to a mixed
perturbation both in $O(V^2)$ and in $O(V) \cdot O( S / B_{12})$. In
fact from Eq. (4.38) we can rederive, to this order, the Schwarzian  given in
Eq. (4.20).

\section{ Discussion}

We have shown here that the BCV gauge can be extended to the case of spinning
particles in 2+1-Gravity, in the region external to some ``CTC
horizons'' that occur around the particles themselves.

Let us note that the existence of the BCV gauge is in general related
to the lack of CTC's. In fact, in our conformal Coulomb gauge the
proper time element is of the form \cite{a11}

\beq ds^2 = \alpha^2 dt^2 - e^{2\phi} {| dz - \beta dt |}^2 \eeq

and, if some istantaneous motion is possible ( $dt = 0$ ) the proper
time

\beq ds^2 = - e^{2\phi} {|dz|}^2  \ \ \ \ \ \ \ \ \ \ \ \ \ \ \ ( dt = 0 )\eeq 
is necessarily spacelike, unless

\beq e^{2\phi} \simeq |g| \simeq {( 1 - {| f |}^2 )}^2 = 0 , \eeq
a case in which it becomes light-like.

Therefore, the (closed) curves defined by $|f| = 1$ are, at the same
time, the boundary for the existence of CTC's, and also for the existence
of the gauge itself, because the metric determinant vanishes. This is
not surprising, because our gauge allows the definition of a
single-valued global time, which is expected to be impossible in the
case of CTC's.

We have provided explicit solutions for the mapping function in
various cases. Firstly for $N$ spinning particles at rest, a case in which
the Schwarzian shows $3N-1$ singularities, and in particular for
$N=2$, a case in which we have also given a closed form for the metric
(Eq. 3.22 ).

Secondly we have also described metric and motion for two spinning
particles, in the quasi-static and the non-relativistic cases. In
particular, we have shown that it is possible to determine the motion
of the particle sites $\xi_i (t)$ ( as singularities of the Schwarzian )
by imposing the $O(2,1)$ monodromies on the exterior solution (
Sec. 4.2 and Appendix A).

Actually, since the Minkowskian coordinates are sums of analytic and 
antianalytic functions of the regular ones in the BCV gauge, it turns
out that each one of them can be continued in the interior
region. This suggests that perhaps the gauge can be extended to the interior
region by relaxing the conformal condition.

We feel however that the clear delimitation of CTC horizons, with
sizes related to the spins, is actually a quite physical feature of
our gauge and points in the direction that a pointlike spin in (2+1)-Gravity
is not really a self consistent concept.

{\bf Acknowledgments }

We are happy to thank Alessandro Bellini for his collaboration in the
early stages of this work, and for valuable discussions. 

This work is supported in part by M.U.R.S.T. , Italy.

\appendix
\section{Analytic continuation of monodromy conditions}

We schetch here the proof that the monodromy conditions can be solved
by the analytic continuation in the mass range of the integrals
defining $f(z)$ and the $X$-mapping. 

Let us first discuss the question of defining the polydromy of $f(z)$
. Since in the physical mass range $f(z)$ has a divergent behaviour around the
particles, its behaviour for $ z \rightarrow \zeta_i$
 cannot be used directly to test the monodromy conditions

\beq f(z) \rightarrow \frac{a}{a*} f(z) + \frac{b}{a*} \eeq
For example, the quasi-static two-particle solution for $f(z)$ is
given by the expression

\beq f_{(\xi_0)} = \int^z_{\xi_0} f' (t) dt = K \int^z_{\xi_0} \ 
( t - \eta_1 ) ( t - \eta_2 ) {( t - \xi_1 )}^{\mu_1 -2} {( \xi_2 - t
)}^{\mu_2 -2} dt \eeq
which diverges for $t \rightarrow \xi_i$ in the physical range $ 0 <
\mu_i < 1$. In order to take the limit $\xi_0 \rightarrow \xi_i$ and
to verify the monodromy conditions, this integral can be defined by
the expression

\beq f_{(\xi_0)} (z) \equiv - \frac{1}{(1-\mu_1)(1-\mu_2)} 
\frac{\partial^2}{\partial \xi_1 \partial \xi_2} \ K \int^z_{\xi_0} \ 
(t-\eta_1) (t-\eta_2) {(t-\xi_1)}^{\mu_1-1} {(\xi_2 -t)}^{\mu-2-1} dt
\eeq
which is now convergent at the endpoints $\xi_i$.

We can then decompose it as follows:

\beq f_{(\xi_0)}(z) = f_{(\xi_0)} (\xi_1) + f_{(\xi_1)} (z) \eeq

Imposing the monodromy conditions (A1) we obtain

\beq f_{(\xi_0)} (\xi_i)  = - \frac{{\overline V}_i}{2} \eeq
which implies

\beq \frac{{\overline V}_{12}}{2} = - \frac{K}{(1-\mu_1)(1-\mu_2)}
\frac{\partial^2}{\partial \xi_1 \partial \xi_2} \int^{\xi_2}_{\xi_1}
\ dt ( t-\eta_1 ) ( t-\eta_2 ) {( t-\xi_1 )}^{\mu_1-1} {( \xi_2-t
)}^{\mu-2-1} |_{\eta_1, \eta_2 {\rm \ const. }} \eeq

Evaluating this expression in the particular case $\xi_1 = 0$, $\xi_2 =
1$, and substituting the relations (3.13) we obtain:

\begin{eqnarray} \frac{{\overline V}_{12}}{2} & = & - K \frac{\Gamma
(\mu_1 ) \Gamma(\mu_2) }{ \Gamma (\mu_1+\mu_2) } \left[ 1 - \sigma_1 
\frac{(1-\mu_1-\mu_2)}{1-\mu_1} - \sigma_2
\frac{(1-\mu_1-\mu_2)}{1-\mu_2} \right] \nonumber \\
& = & - K B ( \mu_1 , \mu_2 ) \left( 1 - \frac{s_1+s_2}{J} \right) 
\end{eqnarray}

which coincides with the result in Eq. (4.8), obtained by analytic continuation
in the parameters $\mu_i$ from $2>\mu_i>1$.

Analogously, the $X$- mapping for two particles  is defined by an
integral which is divergent at the particle sites. However we can
measure the distance $X_1-X_2$ in
the $X$-coordinates by looking at the center of rotation of the
monodromy around the particle sites.

Consider for example the two-particle static case, where the $Z$-mapping
can be defined from a generic point $\zeta_0$:

\beq Z - Z_{\zeta_0} = \frac{R}{K} \int^\zeta_{\zeta_0} \ dt \
t^{-\mu_1} {(1-t)}^{-\mu_2} + {\overline R}{\overline K}
\ \int^{\overline \zeta}_{{\overline \zeta}_0} \ dt \ t^{\mu_1-2} \
{(1-t)}^{\mu_2-2} \ {(t-\tau)}^2 \eeq

Integrating by part around particle 1, we can distinguish a divergent term
which has simple polydromy properties from a remaining series of convergent
integrals:
\begin{eqnarray} Z - Z_{\zeta_0} & = & \frac{R}{K} \int^\zeta_{\zeta_0} \ dt \
t^{-\mu_1} {(1-t)}^{-\mu_2} \ + \ {\overline R}{\overline K} 
\frac{{\overline \zeta}^{\mu_1-1}}{\mu_1-1} {({\overline \zeta } -
1)}^{\mu_2-2} {( {\overline \zeta} - {\overline \eta} )}^2 \nonumber \\
& - & \frac{{\overline R}{\overline K}}{\mu_1-1} 
{{\overline \zeta}_0}^{\mu_1-1} {({\overline \zeta}_0 - 1)}^{\mu_2-2}
{({\overline \zeta}_0 - {\overline \eta})}^2 
- \frac{{\overline R}{\overline K}}{\mu_1-1} 
\int^{\overline \zeta}_{{\overline \zeta}_0} \ dt \ t^{\mu_1-1}
\frac{\partial}{\partial t} ( {(1-t)}^{\mu_2-2} {(t-\eta)}^2 ) 
\end{eqnarray}

Repeating the same reasoning for particle $2$  we obtain the distance  $Z_{12}$
 
\begin{eqnarray} Z_1 - Z_2 & = & - \frac{R}{K} \int^1_0 \ dt t^{-\mu-1} 
{( 1-t )}^{-\mu_2} - {\overline R}{\overline K} {( {\overline \zeta}_0 - 
{\overline \eta}_0)}^2 \left[ \frac{{\overline \zeta}_0^{\mu_1-1} 
{( 1-{\overline \zeta}_0 )}^{\mu_2-2}}{\mu_1-1} - \right. \nonumber \\
 & - &  \left. \frac{{\overline \zeta}_0^{\mu_1-2} 
{( 1-{\overline \zeta}_0 )}^{\mu_2-1}}{\mu_2-1} - \right] + 
\frac{{\overline R}{\overline K}}{\mu_2-1} \int^1_{{\overline
\zeta}_0} \ dz \
{(1-z)}^{\mu_2-1} \frac{\partial}{\partial z} {( z^{\mu-1-2}
{(z - \overline\eta)}^2 )} - \nonumber \\
& - & \frac{{\overline R}{\overline K}}{\mu_1-1} \int^0_{{\overline \zeta}_0}
\ dz \ z^{\mu_1-1} \frac{\partial}{\partial z} ( {( 1-z)}^{\mu_2-2} 
{( z - \overline\eta )}^2 ) \end{eqnarray}

This difference is a finite measure of the distance between the particles
in Minkowskian coordinates. It seems to depend on the choice of the 
arbitrary point $\xi_0$ but in fact it is independent , because of the
relation:

\beq \frac{\partial}{\partial {\overline \zeta}_0} ( Z_1 - Z_2 ) = 0 \eeq

In virtue of this property, the complete evaluation coincides with the
analytic continuation of the integral in Eq. (4.25) from a range where the
mass parameters give a finite result.

\section{ Perturbative expansion of the Schwarzian Derivative }

Given the Schwarzian derivative as

\beq \left( \frac{f''}{f'} \right)' - \frac{1}{2} { \left( \frac{f''}{f'} \right)}^2
= 2 q(z) = \frac{1}{2} \left( \frac{1-\mu^2_i}{{(z-z_i)}^2} + 
\frac{2\beta_i}{z-z_i} \right) \eeq

we can define the lowest order $2q_0 (z)$, given by the choice of the 
parameter $\mu_\infty = \sum_i \mu_i - 1$, which defines a function
$f_{(0)}$. In general, by taking as development parameter 
\beq \delta = \frac{{\cal M}}{2\pi} - \sum_i \mu_i \eeq
the correction term in the Schwarzian at first order in $\delta$ is given by
a sum of poles:

\beq 2 q_1 (z) = \frac{\beta^{(1)}_i - \beta^{(0)}_i }{z-z_i} \eeq
which defines a first order correction to $f(z)$ given by:

\begin{eqnarray}
f'_{(1)} (z) & = & f'_{(0)} (z) \cdot g (z) \nonumber \\
\frac{\partial}{\partial z} \left( \frac{g'}{f'_{(0)}} \right) & = &
2 \frac{q_1 (z)}{f'_{(0)}} \end{eqnarray}
This equation can be easily solved to give the general first-order
solution
\beq g(z) = c_1 + c_2 f_{(0)} + \int^z_{\xi_1} dt f'_{(0)} 
\int^t_{\xi_1} \frac{2 q_1 (w)}{f'_{(0)} (w)} dw \eeq
The constants $c_1$, $c_2$ and the normalization of $f_{(0)}$ have to be
chosen by imposing the monodromy conditions for $f$ which gives, for example,
$c_2=-V_1$. This representation of the solution to the Schwarzian contains 
explicit logarithmic terms around the apparent singularities which
cancel only in the case that the residues $\beta_i^{(1)}$'s satisfy the 
non-logarithmic conditions (4.2).
 
In that case, a simple integration by parts allows to eliminate the poles
in the apparent singularities $\eta_i$ inside the integral (B.5)
reproducing the results (4.17), (4.38) given in the text.




\begin{thebibliography}{999}

\bibitem{a1} A. Staruszkiewicz, Acta Phys. Polonica {\bf 24} (1963) 735.
\bibitem{a2} S. Deser, R. Jackiw and G. 't Hooft, Ann. Phys. (N.Y.) 
{\bf 152} (1984) 220.
\bibitem{a3} J.R. Gott and M. Alpert, Gen. Relativ. Gravitation {\bf 16}
(1984) 243; J.R. Gott, Astrophys. J. {\bf 288} (1985) 422.
\bibitem{a4} G. 't Hooft, Comm. Math. Phys. {\bf 117} (1988) 685.
\bibitem{a5} S. Deser and R. Jackiw, Comm. Math. Phys. {\bf 118} (1988) 495.
\bibitem{a6} P. de Sousa Gerbert and R. Jackiw, Comm. Math. Phys. {\bf
124} (1989) 229; P. de Sousa Gerbert, Nucl. Phys. {\bf B 346} (1990) 440;
G. Clement, Int. J. Theor. Phys. {\bf 24} (1985) 267.
\bibitem{a7} E. Witten, Nucl. Phys. {\bf B311} (1988) 46, {\bf B323}
(1989) 113; A. Achucarro and P. K. Townsend, Phys. Lett. {\bf B180}
(1986) 89, {\bf B229} (1989) 383.
\bibitem{a8} S. Carlip, Nucl. Phys. {\bf B324} (1989) 106; V. Moncrief, J. 
Math. Phys. {\bf 30} (1989) 2907.
\bibitem{a9} A. Cappelli, M. Ciafaloni and P. Valtancoli, Nucl. Phys. {\bf 
B369} (1992) 669; Phys. Lett. {\bf B273} (1991) 431.
\bibitem{a10} G. Grignani and G. Nardelli, Phys. Lett. {\bf B264} (1991) 45;
Nucl. Phys. {\bf B370} (1992) 491.  
\bibitem{a11} A. Bellini and P. Valtancoli, Phys. Lett. {\bf B348}
(1995) 44; A. Bellini, M. Ciafaloni and P. Valtancoli, 
Nucl. Phys. {\bf B454} (1995) 449.
\bibitem{a12} A. Bellini, M. Ciafaloni and P. Valtancoli,
Phys. Lett. {\bf B357} (1995) 532 ; A. Bellini, M. Ciafaloni and
P. Valtancoli, Nucl. Phys. {\bf B462} (1996) 453.
\bibitem{a13} M. Welling, Class. and Quant. Grav. {\bf 13} (1996), 653. 
\bibitem{a14} J. R. Gott, Phys. Rev. Lett. {\bf 66} (1991) 1126;
S. Deser, R. Jackiw and 't Hooft, Phys. Rev. Lett. {\bf 68} (1992) 2647;
P. Menotti and D. Seminara, Phys. Lett. {\bf B301} (1992) 25.
\bibitem{a15} M. Welling, Ref.(13).
\bibitem{a16} See, e.g., M. Yoshida ``Fuchsian Differential
Equations'', Max-Planck-Institut f\"ur Mathematik, Bonn (1987), Fried.
Vieweg \& Sohn Editor.
\bibitem{a17} R. Arnowitt, S. Deser and C.W. Misner, in
L. Witten (ed.) ``{\it Gravitation: an introduction to current
research}'', Wiley, New York (1962).
\bibitem{a18} See, e.g., P. Ginsparg and G. Moore, Lectures given at
the TASI summer school, Boulder CO, published in Boulder TASI 92, 227.
\bibitem{a19} R. Fuchs, Mat. Annalem {\bf 63} (1907), 301.
\bibitem{a20} A.J.Hanson, T. Regge and C. Teitelboim  - ``{\it
Constrained Hamiltonian Systems }'', Accademia Nazionale dei Lincei (1976).

\end{thebibliography}
\end{document}